\documentclass{appolb}
\usepackage{epsfig}
\usepackage[T1]{fontenc}
\usepackage{amsmath,amssymb}
\usepackage{url}

\newcommand{\reffig}[1]{Fig.~\ref{#1}}

\newcommand{\refcite}[1]{Ref.~\cite{#1}}

\newcommand{\dd}{\mathrm{d}}
\newcommand{\kz}{k^0}
\newcommand{\pz}{p^0}
\newcommand{\subtot}{\text{tot}}

\newcommand{\subcoll}{\text{coll}}
\newcommand{\res}[3]{#1$_{#2}$(#3)}

\newcommand{\myvec}[1]{\boldsymbol{\mathrm{#1}}} 
\newcommand{\myunit}[1]{\mbox{$\,\text{#1}$}}
\newcommand{\GeV}{\myunit{GeV}}

\begin{document}

\title{Hadronic transport approach to neutrino nucleus scattering: the Giessen BUU model
  and its validation
  \thanks{Presented by T. Leitner at the 45th Winter School in Theoretical Physics
    ``Neutrino Interactions: from Theory to Monte Carlo Simulations'', L\k{a}dek-Zdr\'oj,
    Poland, February 2--11, 2009.}}
\author{T. Leitner, O. Buss, U. Mosel \address{Institut f\"ur Theoretische Physik,
      Universit\"at Giessen, Germany} }
\maketitle
\begin{abstract}
    We present the GiBUU model for neutrino nucleus scattering ($\nu$GiBUU): assuming
    impulse approximation, this reaction is treated as a two step process. In the initial
    state step, the neutrinos interact with bound nucleons. In the final state step, the
    outgoing particles of the initial reaction are propagated through the nucleus and
    undergo final state interactions. In this contribution, we focus on the validation of
    the initial and final state interaction treatment in GiBUU using experimental data for
    pion-nucleus, photon-nucleus and electron-nucleus scattering.
\end{abstract}

\section{Introduction}

Present day long baseline experiments as MiniBooNE and K2K aim at a precise measurement of
neutrino oscillations. The target material of most modern detectors consists of heavy
nuclei such as carbon, oxygen and iron. To interpret their data, the experiments have to
rely on Monte Carlo event generator predictions for the final state interactions (FSI) in
the target nucleus. For an extraction of electroweak parameters from such experiments it
is, therefore, important to know the expected accuracy of these Monte Carlo analyses.
While most event generators are similar in their treatment of the initial neutrino nucleon
interaction, they differ significantly in their treatment of FSI (more details on the
different models can be found in these proceedings). In addition, in most event
generators, initial state and final state interactions are considered independently.

The Giessen Boltzmann-Uehling-Uhlenbeck~(GiBUU) framework models the full space-time
evolution of the phase space densities of all relevant particle species during a nuclear
reaction within a consistent treatment of the initial vertex and the final state
processes; and we emphasize that these should not be treated separately. This space-time
evolution is determined by the so-called BUU equations, which describe the propagation of
the particles in their potentials and also the collisions between them - resonances are
treated explicitly. A major strength of the GiBUU model is that it has been applied to
many different reactions from heavy ion collisions to pion and electron induced processes
\cite{Teis:1996kx,Lehr:1999zr,Buss:2006vh,Leitner:2008ue} (see \refcite{gibuu} for more
applications). The comparison with data for the reactions mentioned allows to make
estimates for the expected accuracy in neutrino-induced reactions. Unlike most Monte Carlo
event generators, we do not tune any specific input (like for example pion absorption
cross sections) to describe a specific reaction channel (like for example neutrino induced
pion production). To the contrary, we include as much physics as possible and are thus in
a position to explain simultaneously a wide range of very different reactions.  In our
understanding, these are the main points where we differ from event generators like
NUANCE, NEUT and others.

In this contribution, we first introduce the GiBUU model focussing on issues relevant for
neutrino nucleus scattering. Then, we give examples for the applicability of our model and
present results for pion-, photon- and electron-induced reactions which are confronted
with experimental data. After this model validation, we give predictions for
neutrino-induced pion production and nucleon knock-out.

\section{Neutrino nucleus scattering in the GiBUU model}
In the GiBUU model, neutrino nucleus scattering is treated as a two step process. In the
initial state step, the neutrinos interact with nucleons embedded in the nuclear medium as
explained in detail in \refcite{Leitner:2008ue}. The reaction products are, in the final
state step, transported out of the nucleus using a hadronic transport approach
\cite{gibuu}. We will give a brief introduction to both steps in the following.

We treat the nucleus as a local Fermi gas of nucleons bound in a mean-field potential and
obtain for the total neutrino-induced cross section on nuclei
\begin{equation}
  \dd \sigma (\nu_\ell A \to \ell' X') =  \int \dd^3 r \int^{p_F(r)} \frac{\dd^3 p }{(2\pi)^3}  \frac{k \cdot p}{\kz \pz}  \dd \sigma_\subtot^{\text{med}} (\nu_\ell N \to \ell' X) M_{X'}.
\end{equation}
In the latter equation, $k$ is the four-vector of the the neutrino, $p$ the one of the
bound nucleon, and $p^{n,p}_F(r)=(3\pi^2\rho^{n,p}(r))^{1/3}$ denotes the local Fermi
momentum depending on the nuclear density. $M_{X'}$ is the multiplicity of the final state
$X'$ given an initial state X. This mapping $X \to X'$ is determined by the GiBUU
transport simulation described below. The term $\dd \sigma_\subtot^{\text{med}}$ stands
for the total cross section on nucleons including nuclear medium corrections.

In the region of intermediate lepton beam energies ($E_{\text{beam}} \sim 0.5-2 \GeV$),
the total lepton nucleon cross section $\dd \sigma_\subtot$ contains contributions from
quasielastic scattering (QE: $ \ell N \to \ell' N'$), resonance excitation (R: $ \ell N
\to \ell' R$) and direct, i.e., non-resonant, single-pion production (BG: $ \ell N \to
\ell' \pi N'$) treated in our description as background.  The single-pion region is
strongly dominated by the excitation of the $\Delta$ resonance \res{P}{33}{1232}, however,
we include in addition 12 $N^*$ and $\Delta$ resonances with invariant masses less than 2
GeV. The vector parts of the single contributions are obtained from recent analyses of
electron scattering cross sections.  The axial couplings are obtained from PCAC (partial
conservation of the axial current), and, wherever possible, we use neutrino nucleon
scattering data as input.
The neutrino nucleon cross sections are modified in the nuclear medium, i.e., $\dd
\sigma_\subtot \to \dd \sigma_\subtot^{\text{med}}$. Bound nucleons are treated within a
local Thomas-Fermi approximation which naturally includes Pauli blocking. The nucleons are
bound in a mean-field potential depending on density and momentum which we account for by
evaluating the above cross sections with full in-medium kinematics.  We further consider
the collisional broadening of the final state particles within the low-density
approximation $\Gamma_\text{coll} = \rho \sigma v$ obtained in a consistent way from the
GiBUU cross sections (see below).  Details of our model for the elementary vertex and the
corresponding medium-modifications can be found in \refcite{Leitner:2008ue}.

Once produced inside the nucleus, the particles propagate through the nucleus undergoing
final state interactions (FSI) which are simulated with the coupled-channel semi-classical
GiBUU transport model \cite{gibuu}\footnote{The GiBUU code is available for download on
  our website~\cite{gibuu}.}.  Originally developed to describe heavy-ion collisions, it
has been extended to describe the reactions of pions, photons, electrons, and neutrinos
with nuclei~\cite{Lehr:1999zr,Buss:2006vh,Leitner:2008ue}.

This model is based on the BUU equation which describes the space-time evolution of a
many-particle system in a mean-field potential. For particles of species $i$, it is given
by
\begin{equation}
\left({\partial_t}+\myvec\nabla_p H\cdot\myvec\nabla_r  -\myvec\nabla_r H\cdot\myvec\nabla_p\right)f_i(\myvec{r},p, t) =   I_\subcoll [f_i,f_N,f_\pi,f_{\Delta},...],
\end{equation}
where the phase space density $f_i(\myvec{r},p,t)$ depends on time $t$, coordinates
$\myvec{r}$ and the four-momentum $p$.  $H$ is the relativistic Hamiltonian of a particle
of mass $M$ in a scalar potential $U$ given by $H=\sqrt{\left[ M + U(\myvec{r},p
    )\right]^2 + \myvec{p}^{\,2} }$. The scalar potential $U$ usually depends both on
four-momentum and on the nuclear density.
The collision term $I_\subcoll$ accounts for changes (gain and loss) in the phase space
density due to elastic and inelastic collisions between the particles, and also due to
particle decays into other hadrons. The BUU equations for all particle species are thus
coupled through the collision term and also through the potentials in $H$. A
coupled-channel treatment is required to take into account side-feeding into different
channels.  Baryon-meson two-body interactions (e.g., $\pi N \to \pi N$) are dominated by
resonance contributions and a small non-resonant background term; baryon-baryon cross
sections (e.g., $NN \to NN$, $R N \to N N$, $R N \to R' N$, $N N \to \pi NN$) are either
fitted to data or calculated, e.g., in pion exchange models. The three-body channels $\pi
N N \to NN$ and $\Delta N N \to NNN$ are also included.  This complex set of coupled
differential-integral equations is then solved numerically with the GiBUU code.

All particles (also resonances) are propagated in mean-field potentials according to the
BUU equations. Those states acquire medium-modified spectral functions (nucleons and
resonances) and are propagated off-shell.  The medium-modification of the spectral
function is based both on collisional broadening and on the mean-field potentials, both
depending on the particle kinematics as well as on the nuclear density.  With our
off-shell transport we ensure that after leaving the nucleus, vacuum spectral functions
are recovered. Finally, the final state multiplicity $M_{X'}$ is determined by counting
all asymptotic particles $X'$.

We summarize that FSI lead to absorption, charge exchange and redistribution of energy and
momentum, as well as to the production of new particles. Full details of the GiBUU model
are given in \refcite{gibuu} and references therein.

\section{Model validation: pion and electron scattering}

Before applying the above introduced transport model to neutrino-nucleus reactions, we
first evaluate its quality by comparing its predictions for various experimentally
accessible reactions to data. We restrict ourselves to electron- and photon-induced
reactions on nuclei which are in the initial state similar to neutrino nucleus reactions,
and to pion-induced reactions which test our description of the $\pi N \Delta$ dynamics in
nuclei directly.

%
\begin{figure}[tbp]
\centering \includegraphics[scale=.72]{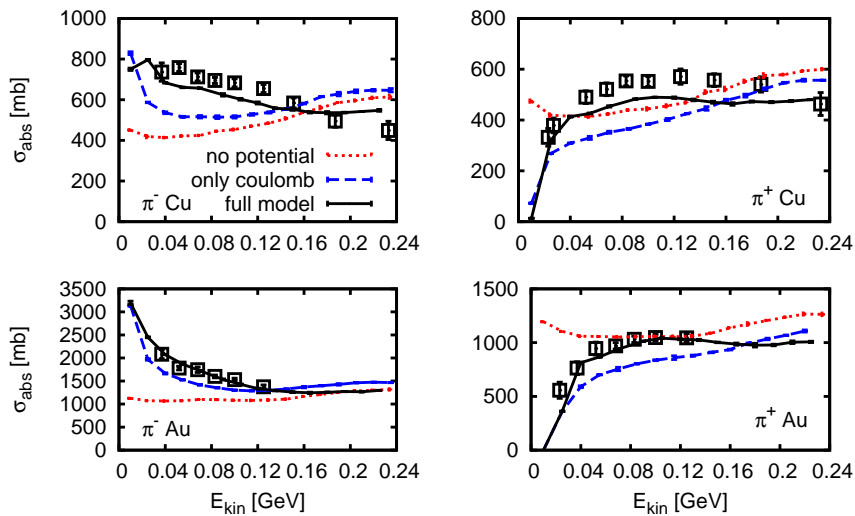}
\caption{Pion absorption on nuclei depending on the choice of potentials for the
pion. The data points are taken from \refcite{nakai}.
\label{fig:pionabs}}
\end{figure}
Let us first focus on pion absorption on nuclei, which directly tests the pion FSI (for
details cf.~\cite{Buss:2006vh}). In \reffig{fig:pionabs} we show calculations for two
different nuclei as a function of the pion energy for both, $\pi^\pm$. Comparing the
curves obtained without any potential (dotted) to those with the Coulomb potential
included (dashed), we see that the Coulomb potential is non-negligible in particular for
heavier nuclei.  We see a reduction of the cross section for the $\pi^{+}$ and a large
increase of the cross section for the $\pi^{-}$ meson. This agrees with the findings of
Nieves et al.~\cite{osetPionicAtoms}, who pointed out the relevance of the Coulomb
potential in their quantum mechanical calculation of absorption and reaction cross
sections.  When one, in addition, includes the hadronic potential for the pion (solid
lines), it adds up with the Coulomb potential to a strongly repulsive potential in the
case of a $\pi^{+}$, while, for the negative pion, at very low energies the two potentials
can even compensate each other (in particular for light nuclei). Overall, we conclude,
that the proper inclusion of the potentials is required to obtain good agreement with the
data.

\begin{figure}[tbp]
\centering \includegraphics[scale=1.0]{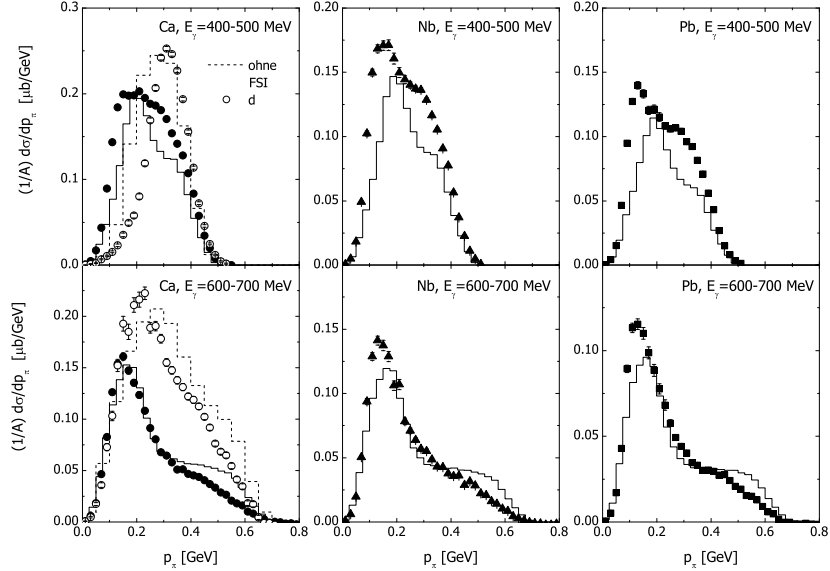}
\caption{Momentum differential cross section for photon induced $\pi^0$ production for
  different nuclei and energies. For Ca also the results without FSI are shown (dashed
  curves) and compared to deuterium data (open circles). Data are from
  \refcite{Krusche:2004uw}. (from \refcite{lehr_phd})
  \label{fig:photo_pionprod}}
\end{figure}

The BUU results for photon induced $\pi^0$ production, which directly test the
vector-interaction part of $\nu$-induced $\pi^0$ production, are shown in
\reffig{fig:photo_pionprod} for Ca, Nb and Pb nuclei for various photon energies as a
function of the pion momentum. Good agreement with the TAPS data \cite{Krusche:2004uw} is
found, in particular the shape is reproduced (for further comparisons see
\refcite{Krusche:2004uw}). Also the results without FSI (dashed lines) and the deuterium
data (open circles) show very similar shapes.  Comparing the dashed with the solid lines
(results without FSI vs.~full calculation), one finds a considerably change of the
spectra.  This shape change is caused by the energy dependence of the pion absorption and
rescattering cross sections. Pions are mainly absorbed via the $\Delta$ resonance, i.e.,
through $\pi N \to \Delta$ followed by $\Delta N \to NN$. This explains the reduction in
the region around $p_\pi=0.3-0.5\GeV$.  In addition, pion elastic scattering $\pi N \to
\pi N$ reshuffles the pions to lower momenta.

\begin{figure}[tbp]
\centering \includegraphics[scale=.75]{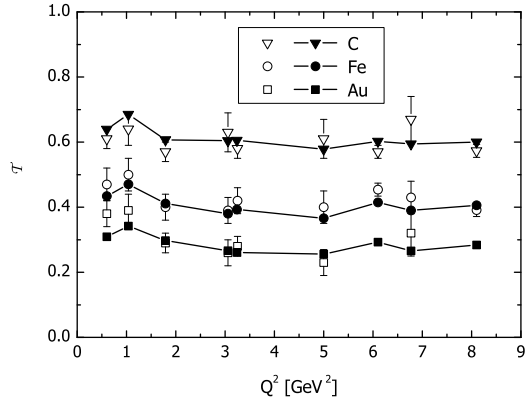}
\caption{Transparency ratio for C, Fe and Pb compared to data (open symbols) from JLab and
  SLAC \cite{O'Neill:1994mg,Abbott:1997bc,Garrow:2001di}. The full symbols connected by
  the lines give the prediction of GiBUU. (from \refcite{lehr_phd})
  \label{fig:transparency}}
\end{figure}

A promising test of the FSI strength is the transparency ratio in $A(e,e' p)$ reactions.
It is defined as the ratio of protons leaving the nucleus with FSI to those without FSI.
While the former quantity can be measured, the latter is purely theoretical and special
assumptions have been made in the experimental analyses
\cite{O'Neill:1994mg,Abbott:1997bc,Garrow:2001di}. \reffig{fig:transparency} shows the
GiBUU result for the transparency ratio for different nuclei as a function of the
four-momentum transfer $Q^2$. The agreement to data is perfect over the wide range of
$Q^2$. For details we refer to \refcite{Lehr:2001an}.

\section{Application to neutrino scattering}
\begin{figure}[tbp]
\includegraphics[scale=.54]{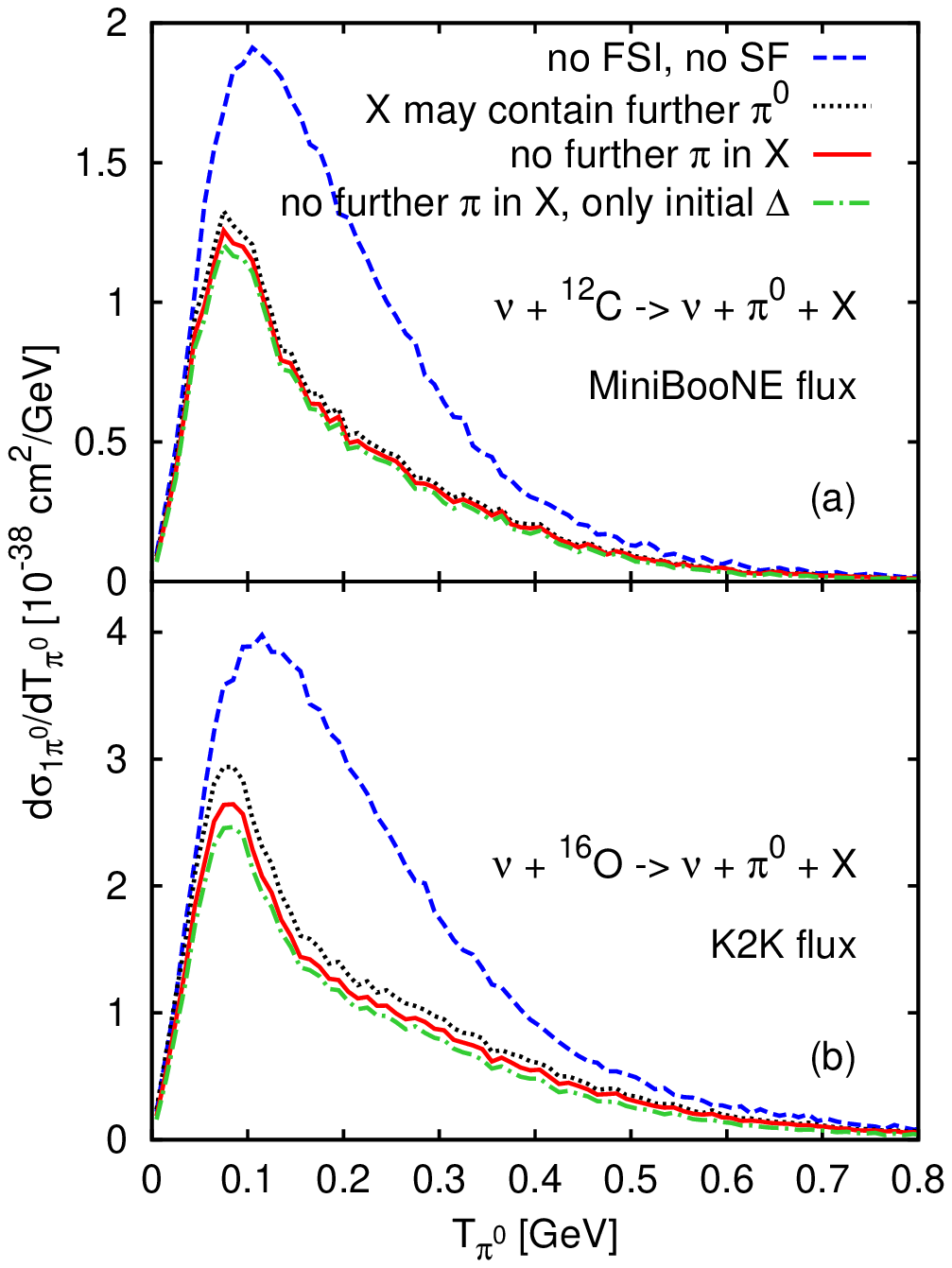} \hspace{0.5cm}
\includegraphics[scale=.54]{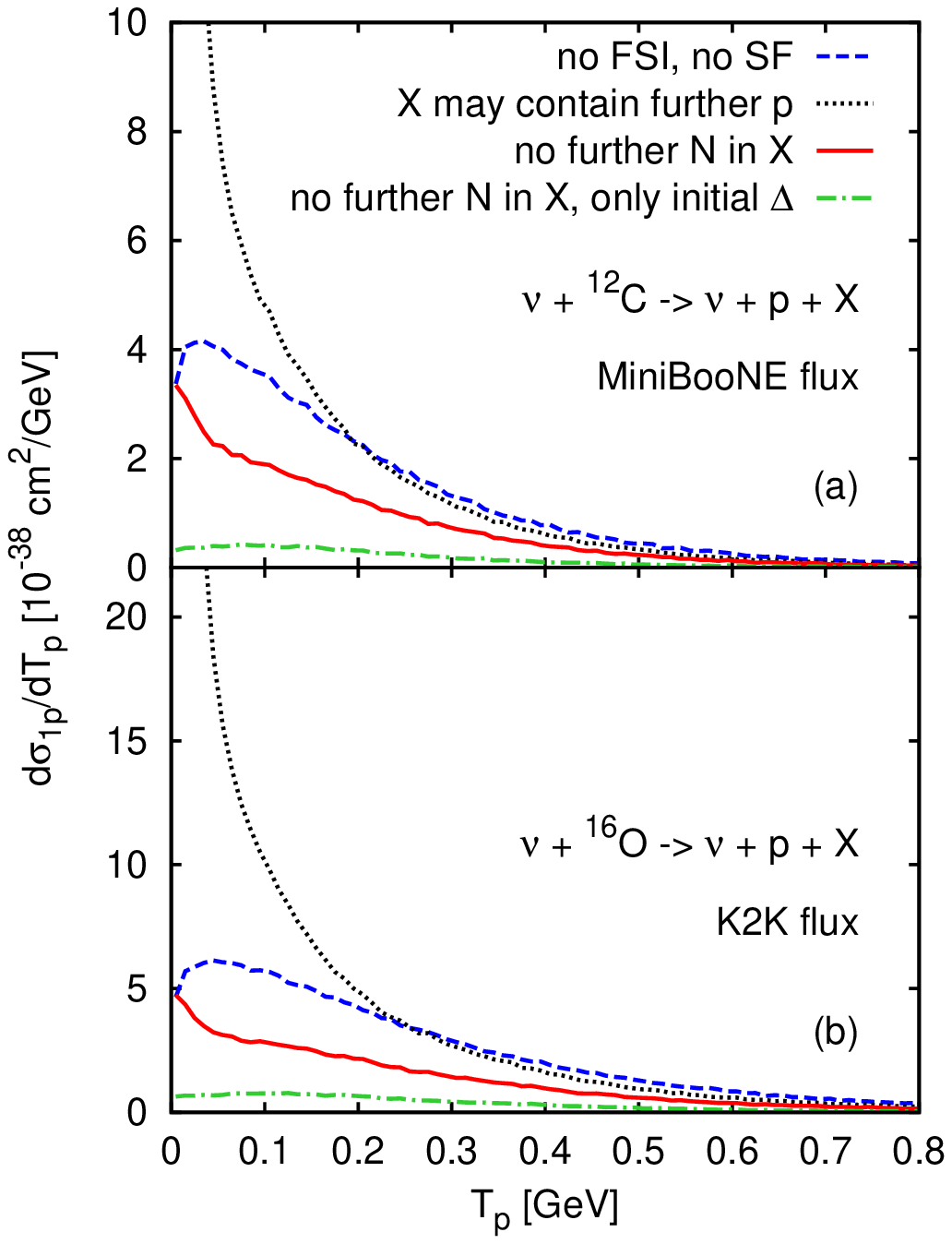}
\caption{Left panel: (a) NC induced $\pi^0$ production on $^{12}$C as a function of
    the pion kinetic energy averaged over the MiniBooNE flux. (b) same on
    $^{16}$O averaged over the K2K flux. Right panel: same for proton knockout.
    \label{fig:MiniBooNE_K2K_NC}}
\end{figure}

To visualize the impact of FSI for neutrino induced reactions, we show in
\reffig{fig:MiniBooNE_K2K_NC} the GiBUU results for NC $\pi^0$ production (left panels)
and proton knockout (right panels) as a function of the corresponding kinetic energy
averaged over the MiniBooNE (top; peaks at 0.7 GeV neutrino energy) and the K2K energy
flux (bottom; peak at 1.2 GeV), respectively. The dashed line does neither include FSI nor
in-medium spectral functions; both are included in all the other lines. The dotted lines
stand for a more inclusive result where the final state may contain more than a
single-$\pi^0$ / single-proton.  This condition is applied in the solid lines where $X$
may not contain any other pions or knocked out nucleons. Finally, the dash-dotted lines
indicate the contribution of the initial $\Delta$ excitation to the solid line.

As for the case of photoproduction discussed before we find a major impact of the FSI on
the pion spectra (comparing dashed with the solid line in left panels).  The vast majority
of the pions come from initial $\Delta$ excitation (dash-dotted line).  Both, the
MiniBooNE and the K2K collaboration, have recently measured NC single-$\pi^0$ momentum
spectra \cite{AguilarArevalo:2008xs,Nakayama:2004dp}, however, their data are only
available as count rates and not yet as cross sections.  A direct and meaningful
comparison to these measurements will be possible when acceptance corrected cross sections
are provided.
Also for proton-knockout, FSI are not negligible as already seen before in the
transparency ratios. The rescattering leads in particular to a large number of knocked out
protons at lower $T_p$ (increase in dotted lines), or, if one looks at it more exclusively
(single-proton knockout), a reduction of flux (dashed vs.~solid lines).

\section{Conclusions}
In this contribution, we have introduced the GiBUU model for neutrino-nucleus scattering
with a focus on the FSI treatment. To validate our model, we have presented results for
photon-, electron- and pion-induced reactions which exhibit similar features as
neutrino-induced processes. We have emphasized that in particular the pion kinetic energy
distributions are very sensitive to a realistic description of the $\pi N \Delta$
dynamics.

We conclude from the successful comparison of the GiBUU calculations for pion, photon and
electron induced reactions to experimental data, that the treatment of initial and final
state interactions is under good control and leads to reliable predictions. In this sense,
the above comparisons serve as a direct benchmark for our neutrino calculations.

We thank L.~Alvarez-Ruso for numerous valuable discussions and all members of the GiBUU
team for cooperation. This work has been supported by the Deutsche Forschungsgemeinschaft.


\end{document}